%% file: ms.tex
\documentclass[twocolumn]{aastex63}
\usepackage{amssymb,graphicx,tikz}
\usepackage{natbib}
\usepackage{multirow}

\let\tablenum\relax
\usepackage{siunitx}

\setcitestyle{notesep={ }} 
\newcommand{\dv}{\ensuremath{\Delta v_{\rm los}}}
\newcommand{\dproj}{\ensuremath{d_{\rm proj}}}

\newcommand{\XMM}{{XMM-\textit{Newton}}}
\newcommand{\RM}{{redMaPPer}}
\newcommand{\arcs}{\ensuremath{^{\prime\prime}}}

\begin{document}

\title{A New Dissociative Galaxy Cluster Merger: RM J150822.0+575515.2}

\shorttitle{A New Galaxy Cluster Merger}

\author[0000-0002-6217-4861]{Rodrigo Stancioli}
\affiliation{Department of Physics and Astronomy, University of California, Davis, CA 
  95616 USA}
\author[0000-0002-0813-5888]{David Wittman}
\affiliation{Department of Physics and Astronomy, University of California, Davis, CA 
  95616 USA}
\author[0000-0002-4462-0709]{Kyle Finner}
\affiliation{IPAC, California Institute of Technology, 1200 E California Blvd., Pasadena, CA 91125, USA}
\author[0009-0007-5074-5595]{Faik Bouhrik}
\affiliation{California Northstate University, 2910 Prospect Park Dr,
  Rancho Cordova, CA 95670 USA}

\correspondingauthor{Rodrigo Stancioli}
\email{rstancioli@ucdavis.edu}


\begin{abstract} 
Galaxy cluster mergers that exhibit clear dissociation between their dark matter, intracluster gas, and stellar components are great laboratories for probing dark matter properties. Mergers that are binary and in the plane of the sky have the additional advantage of being simpler to model, allowing for a better understanding of the merger dynamics. We report the discovery of a galaxy cluster merger with all these characteristics and present a multiwavelength analysis of the system, which was found via a search in the redMaPPer optical cluster catalog. We perform a galaxy redshift survey to confirm the two subclusters are at the same redshift (0.541, with $368\pm519$ km s$^{-1}$ line-of-sight velocity difference between them).
The X-ray morphology shows two surface brightness peaks between the brightest cluster galaxies (BCGs).
We construct weak-lensing mass maps that reveal a mass peak associated with each subcluster. Fitting Navarro-Frenk-White profiles to the lensing data, we find masses of
$M_{\rm 200c}=36\pm11\times10^{13}$ and $38\pm11\times10^{13}$ M$_\odot \; h^{-1}$ for the southern and northern subclusters, respectively.
From the mass maps, we infer that the two mass peaks are separated by $520^{+162}_{-125}$ kpc along the merger axis, whereas the two BCGs are separated by 697 kpc.
We also present deep GMRT 650 MHz data to search for a radio relic or halo and find none.
Using the observed merger parameters, we find analog systems in cosmological n-body simulations and infer that this system is observed between 96 and 236 Myr after pericenter, with the merger axis within $28^{\circ}$ of the plane of the sky.
\end{abstract}

\section{Introduction}\label{sec-intro}

The gravitational interaction between clusters of galaxies may result in large-scale collisions in which two or more clusters plunge toward each other at speeds that can easily exceed $1000 \; \si{km \; s^{-1}}$. As the clusters collide and merge on a timescale of a few Gyr, the energy dissipated is of order $10^{64}\;\si{erg}$, and the total mass of the resulting system can often surpass $10^{15} \; \text{M}_\odot$ \citep{ReinoutRadioReview19}. 
Not surprisingly, therefore, galaxy clusters are a valuable test bed for a myriad of astrophysical phenomena.
One example of this is the influence of mergers on galaxy evolution, as it has been suggested that merging activity can either foster or quench star formation in cluster galaxies \citep{Brodwin2013, Mansheim2017hiz}, as well as increase active galactic nuclei activity \citep{Moravec2020, Sobral2015, MillerOwen2003}. Also of interest are the disturbances in the intracluster medium (ICM) of colliding clusters \citep{Nagai2013}, such as bow shocks and cold fronts \citep{Ghizzardi2010, Markevitch2007}. Such shocks often result in non-thermal, extended radio emissions termed radio relics, where free electrons in the ICM are accelerated to relativistic speeds by the shock and interact with weak magnetic fields to emit synchrotron radiation. The exact mechanism responsible for these relics---and whether they are caused by diffuse shock acceleration alone or by the reacceleration of an existing population of high-energy electrons \citep{kang2021, 2021finner, botteon2020, vanWeeren2017Abell3411, Reinout_toothbrush_2016, vazza2014}---is still not fully understood.

Cluster mergers may also probe dark matter (DM) properties. As the two (or more) subclusters collide, their ICMs exchange momentum and slow down.
The cluster galaxies, on the other hand, are effectively collisionless. Soon after the first pericenter passage, most of the gas is usually found in between the two galaxy-overdense regions. In a cold dark matter (CDM) framework, the DM particles will follow the collisionless behavior of the galaxies and dissociate from the gas. The comparison between X-ray intensity---which traces the ICM morphology---and mass maps derived from weak-lensing (WL) modeling of the Bullet Cluster revealed a significant offset between the positions of the ICM gas and the bulk of the cluster's mass, providing direct evidence of DM \citep{Clowe06}. 

Furthermore, these dissociative mergers are suitable for testing self-interacting DM (SIDM) models \citep{Markevitch04}, as the massive collisions provide plenty of opportunity for DM particles to scatter off each other. The high relative velocity between colliding DM halos means merging clusters are complementary to dwarf galaxies in the investigation of DM properties, especially in view of the possibility that DM self-interaction is velocity-dependent \citep{Sagunski2021, Kaplinghat2016}. 

Further analysis of the Bullet placed an upper bound on the DM self-interaction cross section at $~0.7 \; \si{cm^{2} \; g^{-1}}$ \citep{Randall2008}. Despite ruling out a significant region of the parameter space, this estimate ---along with others derived from more recent analyses of individual clusters \citep{wittman_new_2023, RobertsonBullet2017, Dawson2012musketball}---is still of the same order of magnitude as the cross section of the strong nuclear force. As each observed merger provides only a single snapshot of a gigayear-timescale process, 
placing a stricter upper limit on the SIDM cross section may require discovering an ensemble of merging systems presenting significant dissociation between DM, ICM gas, and stellar components. 

However, measuring the dissociation (or the absence thereof) between DM and galaxies without taking into account the merger stage can lead to erroneous inferences on the DM cross section \citep{Wittman18SIDM, Harvey15}. In an SIDM scenario, the DM halo will lag behind the galaxies soon after pericenter, but the gravitational interaction between cluster components will eventually cause the galaxies to return toward---and through---the DM halo \citep{Kim17}, complicating the analysis. Therefore, it is vital to assemble a set of merging systems that are simple enough to enable accurate modeling, so as to uncover the merger scenario and determine the merging stage. 

In this context, we seek to discover and characterize mergers that are: a) binary, i.e., involving only two subclusters; b) dissociative, with a clear offset between gas and galaxies/DM; and c) near the plane of the sky, which maximizes the possibility of observing such an offset. These features are essential to ensure accurate modeling and improve our understanding of the merger dynamics.

In this paper, we present a new merging system, RM J150822.0+575515.2 (hereafter RMJ1508), that satisfies all these requirements. The discovery method relied on a combination of optical and X-ray data. 
Our starting point is the \RM\ catalog \citep{Rykoff2014}, which contains $\sim25,000$ clusters identified using imaging from the Sloan Digital Sky Survey (SDSS). \RM\ lists the five galaxies that its algorithm considers the most likely to be the brightest cluster galaxy (BCG), along with their respective probabilities of being the BCG. In order to select for bimodality, we require that the most likely BCG has a probability that does not exceed $98\%$ and that the two most likely BCGs have a projected separation of at least $\SI{1}{arcmin}$. We also impose a minimum redMaPPer richness of 120. 
We then search for archival X-ray observations to examine the ICM morphology. Both the X-ray and optical images are manually inspected to confirm that the cluster is in a merging state, with the galaxy overdensities straddling the ICM gas. Our best candidates, with clear signs of bimodality and gas-galaxy dissociation, are then selected for follow-up. 
This is the second cluster discovered with this method for which we publish an in-depth analysis, the first one being Abell 56 \citep{wittman_new_2023}. 
A future paper will address the sample more systematically.

In order to elucidate the merger scenario, we make use of multiwavelength observations, both archival and newly acquired. The paper is organized as follows. Section \ref{sec-overview} presents an overview of the cluster properties based on the current literature. Section \ref{sec-X} analyses archival \XMM\ data to derive the cluster's global X-ray properties. In Section \ref{sec-z}, we present a spectroscopic redshift survey and describe the galaxy kinematics. Section \ref{sec-wl} provides a WL analysis. Section \ref{sec-analogs} derives merger parameters from simulated analogs to reconstruct the merger scenario. The results of our radio observations are reported in Section \ref{sec-GMRT}. Finally, in Section \ref{sec-discussion}, we present our conclusions about the merger scenario.

Throughout the paper, we assume a flat $\Lambda$CDM cosmology with $H_0=69.6 \; \si{km \; s^{-1}}$ and $\Omega_m=0.286$. The cluster is at redshift $z=0.54$, at which $\SI{1}{arcmin}$ corresponds to 385 kpc.

\section{Initial overview}\label{sec-overview}

\begin{figure}
\centerline{
\begin{tikzpicture}
\node at (0,0) {\includegraphics[width=\columnwidth]{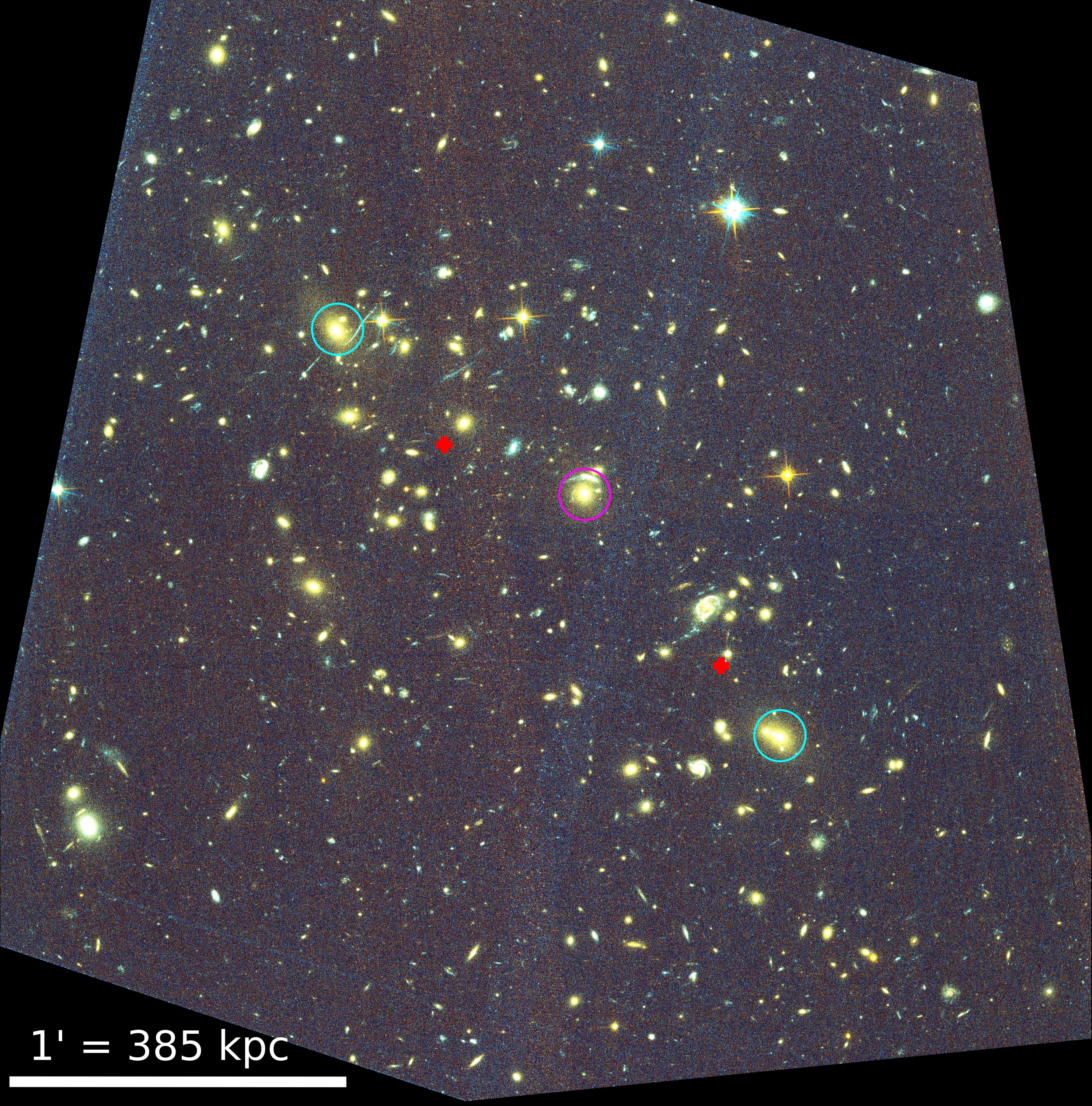}};
\end{tikzpicture}}
\caption{Archival HST/ACS F606W/F814W image of RMJ1508. The NE and SW BCGs are circled in cyan,
and there is a central galaxy (BCG-C, circled in magenta) that is nearly as luminous. The global X-ray peak (red point in the NE) and a local X-ray peak (red point in the SW) are located between the BCGs, suggesting a post-pericenter merger.
}
   \label{fig-ovw}
\end{figure}

In this section, we present a summary of the cluster properties known to the literature. First, we justify our choice of nomenclature. We refer to the cluster by its \RM\ name simply because that was the catalog on which we based our initial search. Historically, however, the cluster was first identified in the catalog produced by \citet{WHL2009}. The updated version of that catalog \citep{WHL2012, WH2015} named this cluster WHL J150811.9+575402 and placed its coordinates on a galaxy that we identify below as the southwestern (SW) subcluster's BCG---as opposed to \RM\, which places the nominal coordinates on the northeastern (NE) BCG.

\textit{BCGs and redshifts.} Using photometric data and archival redshifts of the brightest galaxies in the system, we now provide a brief description of the subcluster configuration.
Figure~\ref{fig-ovw} presents an archival Hubble Space Telescope/Advanced Camera for Surveys (HST/ACS) color (F606W/F814W) 
view of the system. The top two BCG candidates, circled in cyan, are of nearly equal magnitude ($r=19.46$ and $19.45$ for the NE and SW BCGs, respectively, according to SDSS photometry), and are separated by 109$^{\prime\prime}$.
The next brightest galaxy in the cluster ($r=19.90$) is in the center of a gap between the subclusters and is circled in magenta in Figure~\ref{fig-ovw} (hereafter BCG-C).

The NE BCG has a spectroscopic redshift, 0.539273, from the eBOSS
\citep{eboss2020} survey, while the SW BCG has $z=0.53917$ from the
SDSS DR13\footnote{\url{http://www.sdss.org/dr13/data_access/}}. This
yields a line-of-sight velocity difference in the cluster frame,
$\Delta v_{\rm los}=20 \; \si{km \; s^{-1}}$, implying that any relative motion of the
two BCGs is confined to very nearly the plane of the sky.  At this
redshift, the $1.81^\prime$ BCG separation corresponds to 697 kpc.
The BCG projected separation is comparable to that of the Bullet
cluster \citep[720 kpc;][]{BradacBulletLensing2006,Clowe06}, but \dv\
here is much lower than the $616 \; \si{km \; s^{-1}}$ found in the Bullet.

\textit{Richness and mass estimates.}
Although the cluster's mass has not been directly measured previously, many mass estimates can be found in the literature. These estimates are based on scaling relations, and use proxies such as richness, X-ray luminosity, and the Sunyaev-Zel'dovich effect's (SZE) $Y$ parameter. All of those estimates, however, treat the cluster as a single structure instead of considering the presence of two interacting subclusters.

\citet{Rykoff2014} give the
optical richness $\lambda$ (a measure of how many galaxies are in the
cluster, within a certain luminosity range below the BCG) as
152. \cite{RMmassrichnessrelation} calibrated the relation between
WL mass and $\lambda$ (including its scatter), from which we
estimate the mass to be
$M_{\rm 200m}=13.0 ^{+10.0}_{-5.7}\times10^{14}\ h^{-1}$
M$_\odot$ 
\footnote{$M_{\rm \Delta m}$ is the mass enclosed by $R_{\rm \Delta m}$, the radius within which the mean density is $\Delta$ times the mean density of the Universe. Conversely, $M_{\rm \Delta c}$ uses the critical density of the Universe instead of the mean density as the reference level.}. 
\citet{massproxies2017} implemented a system for mass
forecasting with proxies, taking into account various biases, and
found $M_{\rm 200c}=12.37\pm1.90 \times10^{14}$ M$_\odot$ for this system based on its
redMapper richness. They also found $M_{\rm 500c}=9.5\pm 1.0 \times10^{14}$ M$_\odot$
using $Y_{\rm 500}$, a measure of the SZE, as a
proxy. For comparison, \citet{PSZ2} found
$M_{\rm 500c}=5.87^{+0.44}_{-0.43} \times10^{14}$ M$_\odot$ from their
scaling relation based on the same $Y_{\rm 500}$ measurement.

From the optical richness listed in the 2015 WHL catalog and using the calibrated richness from \citet{WH2015}, we estimate the mass to be $M_{\rm 500c}=8.7^{+1.0}_{-0.9}\times10^{14}$ M$_\odot$.

Using a method that combines SZE and X-ray measurements from Planck and ROSAT, \citet{Tarrio2019} estimated the mass at $M_{\rm 500c}=6.15^{+0.69}_{-0.74} \times10^{14}$ M$_\odot$.

\textit{X-ray results from the literature.} We now summarize the cluster's X-ray properties from previous works.
From the scaling relations of \citet{RMscalingrelations2014} one would
expect the X-ray temperature $T_X$ to be around 8 keV with up to
40\% scatter at fixed richness. Because this is a merging cluster, the
X-ray properties may vary from the scaling relations even more than
usual.

\citet{Pratt22} used \XMM\ archival observations (the same observations that we analyze in Section \ref{sec-X}) to estimate the X-ray luminosity and mass of RMJ1508.
They measured the luminosity in the $0.5\text{--}2.0 \; \si{keV}$ range using a core-excised region comprising [0.15--1] $R_{500}$ and found $L_X = 3.87 \pm 0.05 \times 10^{44} \; \si{erg \; s^{-1}}$. The mass was estimated using the $M_{500}-Y_X$ \footnote{The $Y_X$ proxy is the product of the gas mass inside $R_{500}$---estimated from the X-ray surface brightness profile---and the core-excised temperature.}
relation derived by \citet{Arnaud2010}, resulting in $M_{\rm 500c} = 6.15_{-0.24}^{+0.25}\times10^{14} \; \text{M}_\odot$.

Previous analyses of the \XMM\ observations have derived parameters associated with the dynamical state of RMJ1508 in the context of larger samples of clusters \citep{Zhang2023, Bartalucci2019}. In particular, of the four parameters listed in \citet{Yuan2022}, only the concentration ($\log_{10}c = -0.99 \pm 0.04$) is significantly indicative of a disturbed state. The other three parameters estimated by them seem to be inconclusive, which corroborates the virtue of adding optical information to better identify clusters in a merging state, as done in our selection method.

\section{X-ray properties}\label{sec-X}

The \XMM\ Science Archive contains two observations of RMJ1508, performed in 2012 and 2013 (PI Arnaud). The exposure times of both observations for each of the three instruments in the European Photon Imaging Camera (EPIC) are listed in Table \ref{xmm_exposure_time}\footnote{In the 2013 observation (ObsId 0723780501), the MOS exposures comprise two intervals, which were combined in the reduction process using the \texttt{SAS} task \texttt{merge}.}. We performed the data reduction using the \XMM\ Science Analysis System (\texttt{SAS}) version 19.0.0. The observations were cleaned of soft proton flares by selecting good time intervals with less than 0.3 (0.4) counts s$^{-1}$ in the 10-12 keV band in the MOS (pn) detectors. An estimation of the residual soft proton contamination using the script developed by \citet{DeLuca04}\footnote{Available at \url{https://www.cosmos.esa.int/web/xmm-newton/epic-scripts\#flare}.} indicated no noticeable contamination after filtering.

\begin{deluxetable*}{lccclccc}
  \tablecaption{XMM Exposure Times}  \label{xmm_exposure_time}
  \tablecolumns{8}
  \tablehead{\multirow{2}{50pt}{Obs. Id.} & \multicolumn{3}{c}{Total (s)} & \multirow{2}{1pt}{ } & \multicolumn{3}{c}{Filtered (s)} \\
  \cline{2-4} \cline{6-8}
  & \colhead{MOS1} & \colhead{MOS2} & \colhead{pn} & & \colhead{MOS1} & \colhead{MOS2} & \colhead{pn}}
  \startdata
    0693660101 & 14,323 & 14,350 & 12,862 & & 10,897 & 10,902 & 7,030 \\
    0723780501 & 21,621 & 21,628 & 17,746 & & 17,355 & 17,372 & 15,100
    \enddata
\end{deluxetable*}

In order to estimate the global properties of the ICM, we extracted the spectrum of a circular region with a 70 arcsec radius centered at the midpoint between the two subclusters along the merger axis. Point sources were removed using the \texttt{cheese} routine from the \XMM\ Extended Source Analysis Software (\texttt{ESAS}). We accounted for the background emission by using the double-subtraction method described in \cite{Arnaud2002}. We defined the background region as a circle of 100 arcsec radius to the southeast of the cluster without any visibly resolved sources, and we used blank-sky files \citep{Carter2007} to account for the spatial variability of the background components across the detector. All event lists were corrected for vignetting using the \texttt{SAS} task \texttt{evigweight}.

The procedure described above was carried out independently for each one of the three EPIC instruments and two observations, resulting in six spectra. We then performed a simultaneous fit of these spectra to a model consisting of an \texttt{apec} component---corresponding to the thermal bremsstrahlung emission from the cluster---multiplied by a \texttt{phabs} component to account for galactic absorption. The best-fit parameters resulted in a global temperature of 
$T_X = 7.00_{-0.45}^{+0.50} \; \si{keV}$ and a total unabsorbed luminosity of
$L_X = 1.099_{-0.023}^{+0.023} \times 10^{45}\; \si{erg \; s^{-1}}$ in the $0.5-10 \; \si{keV}$ energy range, where the error bars represent the 90\% confidence interval.

An X-ray intensity map (see Section \ref{sec-wl}) was
obtained from the point-source-subtracted, exposure-corrected image in the 0.4-1.25 keV energy range, using only the longer of the two observations (Obs. Id. 0723780501). 
The image was created according to the procedure described in the XMM ESAS Cookbook \citep{ESAS2023} and adaptively smoothed using the \texttt{adapt} routine from \texttt{ESAS}. Further smoothing with a Gaussian kernel ($\sigma= 6.75^{\prime\prime}$) was applied when generating the contours for presentation purposes.

\section{Redshift survey and clustering kinematics}\label{sec-z}

We now analyze spectroscopic and photometric data of cluster members, in order to shed light on the member galaxy distribution and the relative velocities between the subclusters.

\subsection{Photometric Member Distribution}\label{sec-photometric}

\begin{figure}
\centerline{
\begin{tikzpicture}
\node at (0,0) 
{\includegraphics[width=\columnwidth]{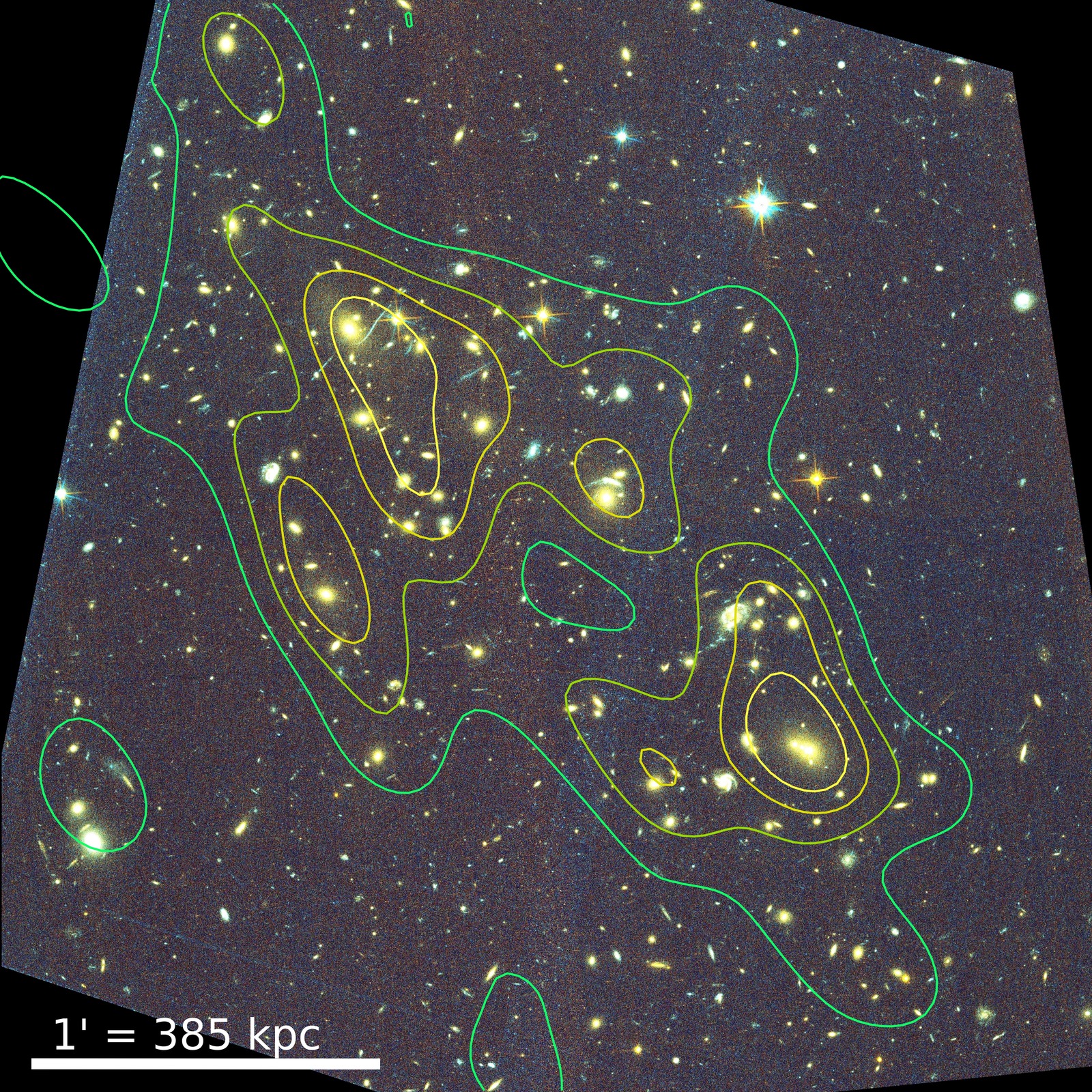}};
\end{tikzpicture}}
\caption{Luminosity-weighted galaxy density contours over the HST/ACS F606W/F814W image. The cluster's bimodality is evidenced by the two overdense regions around the NE and SW BCGs.
Contour levels start at the mean pixel value of the luminosity density map and increase in units of $\sigma$, where $\sigma$ is the standard deviation of pixel values.
}
   \label{fig-galaxy_density}
\end{figure}

In order to reveal the projected distribution of luminous matter in the two subclusters and confirm the system's bimodality, 
we build a photometric catalog of the cluster region using archival HST/ACS imaging (Proposal ID: 14098, PI: Ebeling) in two bands. The F814W image was taken on 2016 February 20, whereas the F606W image was taken on 2016 August 27. Both filters had the same exposure time of 1200 s. After standard instrumental signature removal, we used SExtractor \citep{1996bertin} to detect objects.

A visual inspection of the color-magnitude diagram suggests the red sequence falls in the region $1<\text{F606W}-\text{F814W}<1.5$, so we select galaxies in this color range.
We filter out objects with a high probability of being stars according to SExtractor, as well as objects brighter than the BCG or with an F814W magnitude fainter than 24\footnote{Given our goal of producing a luminosity-weighted density map, excluding faint \textit{bona fide} cluster members has a negligible effect on our results.}. We add to the catalog galaxies that are confirmed to be cluster members by our spectroscopic survey described in Section \ref{subsec-z}, and exclude those confirmed to be in the foreground or background.

We generate a luminosity-weighted galaxy density map by applying the Gaussian kernel density estimator provided by \texttt{scipy.stats.gaussian\_kde} \citep{scipy}
to our cluster member catalog. The results are shown in Figure~\ref{fig-galaxy_density}. The map shows two distinct overdense regions around the SW and NE BCGs, with a clear dearth of luminous galaxies in the center, near BCG-C. The NE density peak appears elongated in the N-S direction, between the NE BCG and a group of galaxies $30"$ to its S. The SW peak is centered on the SW BCG and is slightly elongated toward the N.
The NE subcluster appears to be somewhat more optically rich than the SW subcluster. This is borne out by the BCG probabilities according to \citet{Rykoff2014}: the NE BCG has a probability of 0.89 and the SW BCG has a probability of 0.11 of being the overall BCG. 

\subsection{Redshift Survey}\label{subsec-z}

\textit{Observational setup.} We observed RMJ1508 with the DEIMOS
multiobject spectrograph \citep{FaberDEIMOS} at the W. M. Keck
Observatory on 2022 July 1 (UT). The seeing at the time of observation was roughly 1.2\arcsec. We used the 1200 line mm$^{-1}$ grating for a pixel scale of 0.33 \AA\ pixel$^{-1}$ and a spectral resolution of ${\sim}1$ \AA, which corresponds to $\sim30 \; \si{km \; s^{-1}}$ in the observed frame. The observed wavelength range was $\approx$ 5500--8000 \AA. We designed two slit masks, each one containing ${\sim}50$ slits. The exposure time was $3\times15$ minutes on the first mask, and only $2\times6$ minutes on the second mask due to the target setting.

The object selection and slitmask preparation followed the procedure described in \citet{wittman_new_2023}. In essence, we used Pan-STARRS photometric redshifts \citep{PSphotoz2021} to calculate the likelihood of each galaxy in the field being a cluster member, according to the expression
\begin{equation}
  \label{eq:1}
\mathcal{L}\propto\frac{1}{ \sigma_{\rm  PS}}  \exp \frac{(z_{\rm PS}-z_{\rm cl})^2}{2 \sigma_{\rm
  PS}^2},
\end{equation}
where $z_{\rm PS}$ and $\sigma_{\rm  PS}$ are the Pan-STARRS photometric redshift and its corresponding uncertainty, respectively; and $z_{\rm cl}$ is the cluster redshift. The median value of $\sigma_{\rm PS}$ was 0.17, large enough to retain sensitivity to foreground and background structures.
In order to avoid placing too many slits on faint galaxies unlikely to yield spectroscopic redshifts, we multiplied the likelihood by an apparent magnitude weight, $(24-r)$. The final likelihood was then used to define the galaxy priority values passed as an input to the slitmask design software \texttt{dsimulator}. In retrospect, this apparent magnitude weighting also had the effect of upweighting foreground galaxies relative to background galaxies. As a result, our survey is quite sensitive to potential foreground structures.

\textit{Data reduction and redshift extraction.} The data reduction was performed using PypeIt \citep{pypeit:joss_pub,pypeit:zenodo}. In order to obtain an accurate wavelength calibration, we created a customized template for the wavelength solution using the \texttt{pypeit\_identify} script. The wavelength calibration for each slit was compared to the sky emission lines from our science frames as a consistency check. After obtaining the 1-D spectra with PypeIt, we extracted a redshift for each object by cross-correlating its spectrum with a set of templates using a custom Python code that implements many aspects of the approach used by the DEEP2 survey \citep{Deep2:2013}; see \citet{wittman_new_2023} for more details on this software. We were able to extract a secure redshift for 52 (18) galaxies in the first (second) mask, for a total of 70 galaxies, which are listed in Table~\ref{tab-zspec}.

\begin{deluxetable}{llll}
  \tablecaption{Galaxy redshifts}  \label{tab-zspec}
  \tablecolumns{4}
  \tablehead{\colhead{R.A. (deg)} & \colhead{Decl. (deg)} & \colhead{Redshift} & \colhead{Uncertainty}}
  \startdata
\input{ztable.tex}
    \enddata
    \tablecomments{This table is published in its entirety in
  machine-readable format. A portion is shown here for guidance regarding
  its form and content.}
\end{deluxetable}

\begin{figure}
\centerline{\includegraphics[width=\columnwidth]{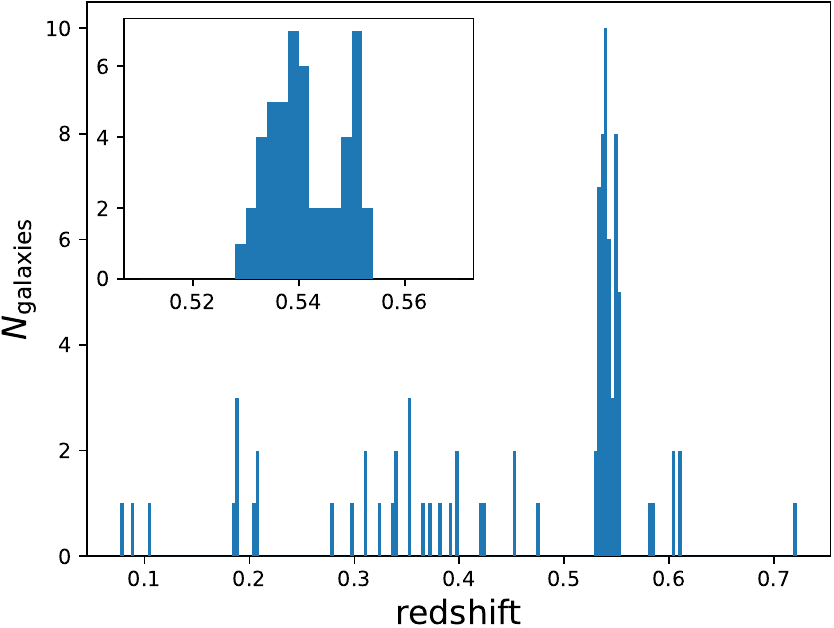}}
\caption{Redshift histogram, with inset showing the redshift interval
 around the cluster.}
\label{fig-zhist}
\end{figure}

\textit{Archival redshifts.} 
In a 10\arcmin\ radius around the cluster's nominal position, we found
16 galaxies in the NASA/IPAC Extragalactic
Database\footnote{\url{http://ned.ipac.caltech.edu}} \citep{ned} with known spectroscopic redshifts.  We found nine additional galaxy redshifts in the eBOSS
\citep{eboss2020}
database\footnote{\url{https://dr17.sdss.org/optical/spectrum/search}} for a
total of 25 archival galaxy redshifts. Of these, seven coincided with
galaxies in our Table~\ref{tab-zspec}; the mean redshift difference
between archival measurements and ours is $(1.3\pm0.9)\times10^{-4}$
with an rms scatter of $2.5\times10^{-4}$; this corresponds to
$25\pm18 \; \si{km \; s^{-1}}$ in the cluster frame with an rms scatter of $48 \; \si{km \; s^{-1}}$. After removing duplicated entries and merging the catalogs, we ended up with redshifts for 88 galaxies. A histogram of the final catalog is shown in Figure~\ref{fig-zhist}.

\subsection{Subclustering and Kinematics}\label{sec-subclustering}

First, we note that the redshift histogram shows no serious foreground or background
cluster candidates that could complicate the interpretation of X-ray or lensing data in this field.
We proceed to analyze the redshift window from
0.52 to 0.56, containing 49 galaxies.  We estimate the mean redshift and
velocity dispersion using the biweight estimator \citep{Beers1990},
with uncertainties obtained via the jackknife method. We find the
systemic redshift to be $0.5410\pm0.0012$ and the velocity dispersion
to be $1398\pm104 \; \si{km \; s^{-1}}$, likely inflated by merger activity. The
histogram suggests that the distribution may be bimodal, so we apply
an Anderson-Darling (AD) test for consistency with a single Gaussian and find a p-value in the range (0.025,0.05), which is suggestive but not definitive.

\begin{figure}
\centerline{\includegraphics[width=\columnwidth]{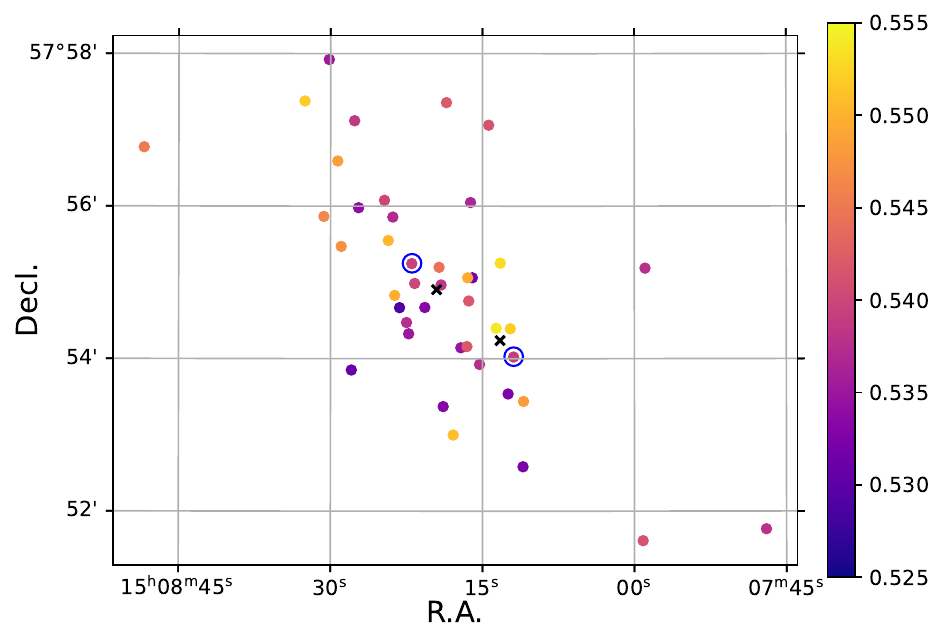}}
\caption{Map of galaxy redshifts in the range $0.52{<}z{<}0.56$. The points circled in blue are the subclusters' BCGs, and the black crosses are the X-ray surface brightness peaks.} There is no apparent correlation between redshift and position, suggesting that any bimodality in the redshift histogram is not due to a radial velocity difference between the two subclusters.
\label{fig-zmap}
\end{figure}

\textit{Redshift bimodality?} A natural explanation for two redshift peaks would be that the two subclusters have some relative velocity along the line of sight.  To visualize any correlation between velocity and sky positions, we present a color-coded map in Figure~\ref{fig-zmap}. There is no apparent correlation between position and velocity, which agrees with the BCG redshifts (Section \ref{sec-intro}) in suggesting that the two subclusters have negligible relative line-of-sight velocity. This raises the question of whether the second redshift peak indicates a third body, or is a statistical fluke.

\begin{figure}
\centerline{\includegraphics[width=\columnwidth]{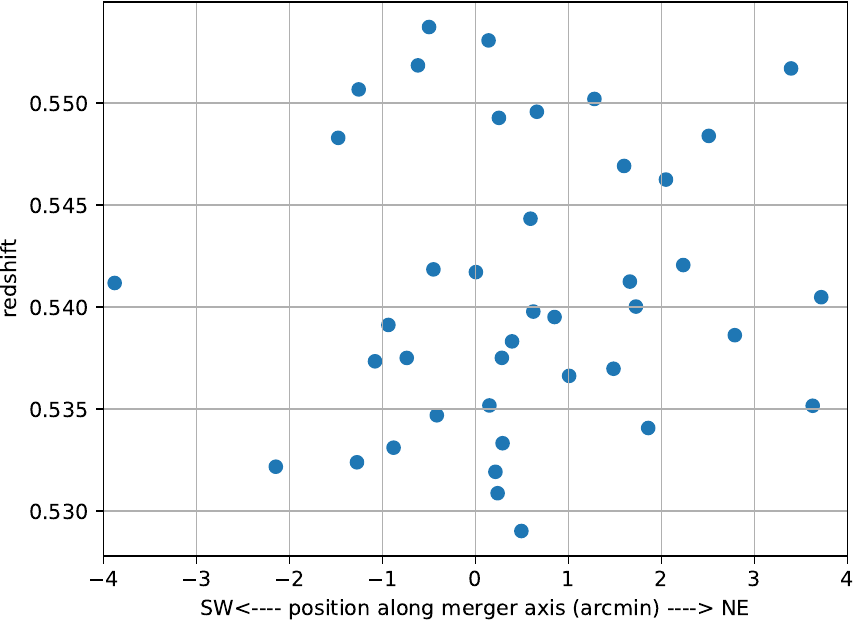}}
\caption{Galaxy redshift vs position along the merger axis. If there is redshift bimodality (with a possible dearth of galaxies around $z=0.545$), it cannot be assigned to either the NE or SW subcluster specifically.}
\label{fig-mergeraxis-z}
\end{figure}

To test this further, Figure~\ref{fig-mergeraxis-z} shows the relationship between galaxy redshift and position along the merger axis, with the NE (SW) subcluster defined as having a positive (negative) position. The magenta-circled galaxy in Figure~\ref{fig-ovw} (BCG-C) appears at the origin in this coordinate system\footnote{See Section \ref{sec-discussion} for more details on the merger axis definition.}.
There does appear to be a dearth of redshifts around $z=0.545$, but this dearth appears in both subclusters, and both subclusters have a similar fraction of galaxies with redshift greater than this. An AD test for consistency with a single Gaussian finds $p>0.15$ ($p=0.10$) for the NE (SW) subcluster. In other words, each subcluster is consistent with a single Gaussian but pooling their galaxies leads to stronger evidence for an additional velocity component. If this bimodality is real, it is difficult to explain with a physical model: the line-of-sight velocity difference between groups at $z{=}0.535$ and $z{=}0.55$ is about $2900 \; \si{km \; s^{-1}}$ in the frame of the cluster, but the substantial projected separation between galaxy groups 
(see Figure~\ref{fig-galaxy_density})
argues strongly against a line-of-sight merger. Hence, we tentatively attribute the dip in the redshift histogram (Figure~\ref{fig-zhist}) to the small spectroscopic sample size rather than a real physical feature. 

\textit{Subcluster relative velocities.} 
In view of the photometric analysis presented in Section \ref{sec-photometric}, which suggests the presence of two subclusters separated in the plane of the sky, we proceed to assign galaxies to subclusters based entirely on their projected positions.
We use the position along the merger axis as visualized in Figure~\ref{fig-mergeraxis-z} to define the NE (SW) subcluster as those galaxies with positive (negative) coordinates, omitting BCG-C, which is difficult to assign to either subcluster. These two subclusters have consistent mean redshifts: $0.5416\pm0.0014$ ($0.5398\pm0.0022$) for NE (SW) based on 31 (17) members, with velocity dispersions of $1420\pm122$ ($1356\pm283$) $\si{km \; s^{-1}}$. The redshift difference between subclusters is $0.0018\pm0.0026$, or $350\pm 507 \; \si{km \; s^{-1}}$ in the frame of the cluster. 

The subcluster velocity dispersions are quite high, even when compared to a sample of merging clusters such as that of \citet{MCCsampleanalysis}. The nature of the putative second redshift peak must be resolved in order to properly interpret this high-velocity dispersion. Note, however, that the velocity dispersion is a strong function of time in a merger, peaking around the time of pericenter \citep{Pinkney96,Takizawa2010}. In Section \ref{sec-analogs} we present evidence that this merger is seen sooner after pericenter than most of the systems in \citet{MCCsampleanalysis}.

\section{WL analysis}\label{sec-wl}

Using the archival HST/ACS F814W imaging described in Section \ref{sec-photometric}, we perform a WL analysis of RMJ1508. For each galaxy identified with SExtractor, we generated a relevant point-spread function (PSF) model following the method of \cite{2007jee} by utilizing their publicly available PSF catalog. Each galaxy was fit with a PSF-convolved Gaussian model parameterized by a pre-PSF complex ellipticity; see \citet{2017finner,2021finner,2023finner} for more details on our ACS WL pipeline. In order to prevent spurious sources such as diffraction spikes around bright stars and poorly fit objects from entering the source catalog, we remove objects with ellipticity greater than 0.8, ellipticity uncertainty greater than 0.3, and intrinsic size (pre-PSF) less than 0.5 pixels. The intrinsic size constraint removes objects that suffer from poor pixel sampling and would introduce systematic biases in shape measurement if not removed \citep{Finner2023_NIRWL}.

Next, we must filter out as many foreground and cluster galaxies as possible from our source catalog and, at the same time, minimize the loss of background sources.
For this step, we use galaxy colors based on the F606W imaging (see Section \ref{sec-photometric} above).
The cluster red sequence follows a linear relation at an F606W -- F814W color of approximately 1.2. We select background galaxies by keeping those with F606W -- F814W color less than 1. In addition, we constrain the F814W magnitude to be fainter than 24. The final source
catalog contains 87.2 galaxies arcmin$^{-2}$ (988 galaxies total). We apply the same color and magnitude cuts to the GOODS-S photometric redshift catalog
\citep{GOODSzphot2013} and find that the contamination by foreground
galaxies is expected to be 6.5\%; this is accounted for when fitting mass models below. Cluster galaxies may contribute additionally to the contamination. However, inspection of a 2-D source density map does not reveal any obvious associations with the known subclusters. 

\begin{figure}
\centerline{
\includegraphics[width=\columnwidth]{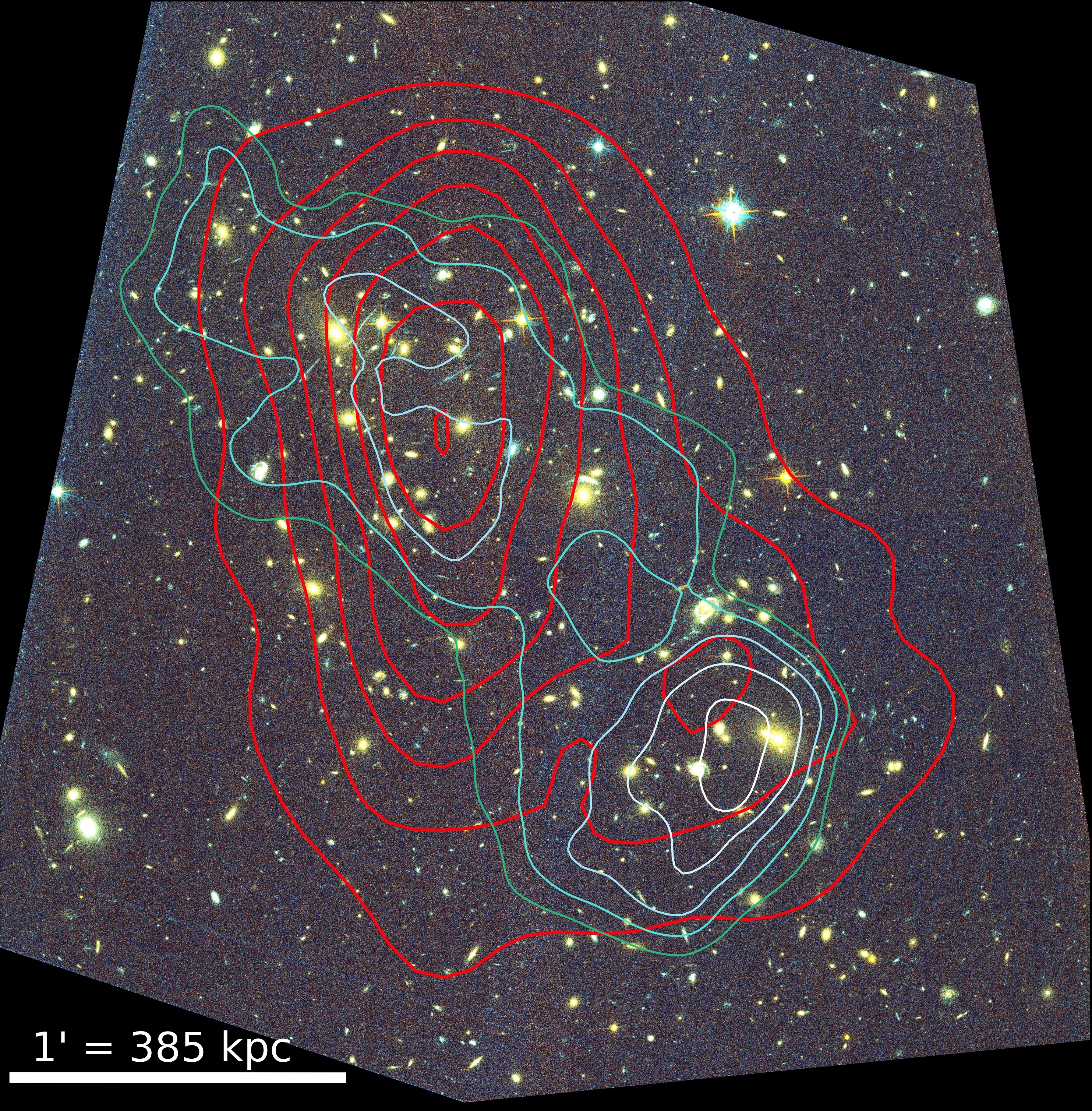}}
\caption{HST/ACS F606W/F814W image with overlaid surface mass density contours from WL. The contour color map goes from green to blue to white in ascending order of 
convergence (smoothed by $8"$) in steps of $0.011$, with the absolute level unknown due to mass-sheet degeneracy. 
(Note that the central closed contour indicates a trough rather than a peak.) The red contours represent \XMM\ surface brightness. The X-ray contour levels start at 14.7 counts s$^{-1} \text{deg}^{-2}$ and increase in steps of 7.4 counts s$^{-1} \text{deg}^{-2}$ in the 0.4-1.25 keV energy range. The SW subcluster has a higher, more well-defined mass peak, whereas the NE subcluster's mass peak is more irregular in shape. The X-ray surface brightness may be described as a NE-SW ridge with a main peak near the NE subcluster and a secondary peak near the SW subcluster, albeit lagging that subcluster if it is outbound.}
\label{fig-wlmap}
\end{figure}

Before fitting mass models, we present a nonparametric reconstruction of the surface mass density (convergence) field using the \texttt{FIATMAP} code \citep{2006wittmanDLS}, which convolves the observed shear field with a kernel of the form
\begin{equation}
r^{-2} (1-\exp({-r^2 \over 2r_i^2})) \exp({-r^2 \over 2r_o^2}),
\end{equation}
where $r_i$ and $r_o$ are inner and outer cutoffs,
respectively. 
Following \citet{wittman_new_2023}, we used $r_i=50 \; \si{arcsec}$ and $r_o=100 \; \si{arcsec}$. The resulting mass map has 1.5 arcsec per pixel.
Figure~\ref{fig-wlmap} shows this
map as a set of contours overlaid on the HST/ACS color image. 
Each galaxy subcluster is associated with a WL peak, with a trough in between. The well-defined SW lensing peak is offset to the SW of the local X-ray peak, supporting the scenario in which this subcluster is traveling SW, outbound after a recent pericenter passage in which ram pressure slowed the gas. The NE lensing peak is less well defined, such that the sign of its offset relative to its local X-ray peak cannot be determined.

We estimate the mass of each subcluster by fitting a two-halo NFW
model with a mass--concentration relation from
\cite{Child18}. We use the \texttt{GalSim} \citep{galsim} library to compute the shear of the two-halo model at the position of each background galaxy, and we set the distance ratio $\beta$ of the sources to the mean distance ratio of sources in the GOODS-S photometric redshift catalog \citep{GOODSzphot2013} meeting the color and magnitude cuts described above ($\bar{\beta}=0.45$ where the bar indicates averaging over sources). Given a lens model, \texttt{GalSim} predicts the reduced shear field $g(\beta)$ for a single source plane at distance ratio $\beta$. However, $g$ is a nonlinear function of $\beta$ such that $\overline{g(\beta)}\ne g(\bar{\beta})$. To correct for this, we use the \citet{Seitz97} approximation \begin{equation}
    \overline{g(\beta)}\approx g(\bar{\beta})\left[1+(\frac{\overline{\beta^2}}{(\bar{\beta})^2}-1)\kappa\right]
\end{equation}
noting that this correction is nontrivial only near the NFW peaks where the convergence $\kappa$ is not negligible.

We used \texttt{scipy.optimize.least\_squares} \citep{scipy} to optimize six parameters: the mass and 2-D positions of each of the two NFW halos.  We fixed the concentration of each halo to the value predicted by the \citet{Child18} mass--concentration relation, after finding that the NE halo does not converge when allowing for scatter in this relation.
We put broad, flat priors on each of the six parameters: (1--300) $\times10^{13}$ M$_\odot$ on the mass and $\pm30^{\prime\prime}$ on the location of each halo. We find the SW (NE) mass to be
$M_{\rm 200c}=36\pm11\times10^{13}$ ($38\pm11\times10^{13}$) M$_\odot \; h^{-1}$; these uncertainties are found by bootstrap resampling the source catalog but are similar to those inferred from the curvature of the likelihood surface. The fitted positions are consistent with the convergence map peaks, regardless of starting position within the bounding box, with uncertainties of about $\pm5^{\prime\prime}$.
Note that \citet{wonki2023} recently found that for merging clusters seen shortly after pericenter, the WL-derived masses tend to be biased high, since the mass profile is disturbed by the merger and deviates from the fitted model profiles. The bias is a function of time since pericenter (TSP) and mass. For RMJ1508, at the most unfavorable TSP, the bias would likely be between 23\% and 41\%. 

Comparing the WL masses with the mass suggested by the proxies listed in Section \ref{sec-overview} is not straightforward given that the references cited there did not treat the system as composed of two clusters. Furthermore, various mass proxies are not expected to agree well in a merging system. Nevertheless, we note that the sum of the WL masses is comparable to the proxy estimates. After accounting for $h^{-1}$, the sum of our WL M$_{\rm 200c}$ masses is $10.6 \pm 2.2\times10^{14}$ M$_\odot$, which is 
consistent with the mass estimates cited in Section \ref{sec-overview} within the error bars.
(Note that the mass inferred from lensing would increase if one draws a larger virial radius based on the combined mass of the two halos in the lensing model, but it is not clear that this procedure provides a more rigorous comparison with the mass proxies.)

As a test of whether a six-parameter model (including mass and 2-D
position parameters for each of the two halos) is justified, we compute the
Bayesian Information Criterion (BIC). Compared to a model with no lens
mass, the two-halo model has a BIC lower by $25$ logarithmic
units, which is considered very strong evidence \citep{KassRaftery95}. Compared to a
model with a single halo in the SW, the two-halo model is again
very strongly preferred ($\Delta\textrm{BIC}=17$). 

We bootstrap-resample the catalog of source galaxies to estimate the uncertainty in the position of each mass peak. We generate 10,000 bootstrap realizations of the mass map and record the global peak position for each one. 
We then use the peak positions as the input to a k-means algorithm (\texttt{sklearn.cluster.KMeans}; \citealt{sklearn}) with the number of clusters fixed at two. The k-means algorithm iteratively assigns each recorded peak to one of the two clusters and calculates the two centroid positions, minimizing the distance between the data points and the centroid within each group. After convergence is obtained, we find that the peak is located in the SW (NE) subcluster in 61\% (39\%) of the bootstrap realizations. By randomly sampling pairs of NE and SW peaks, we estimate the 68\% confidence interval on the distance between peaks. We find that the NE and SW mass peaks are separated by $520^{+162}_{-125}$ kpc, with most of the uncertainty stemming from uncertainty in the position of the northern mass.

The bootstrap-resampling method can also be used to assess the detection significance of each peak. In order to do so, we estimate the noise level at each pixel in the mass map by computing the rms of the pixel value across all 10,000 realizations. We then divide the average mass map by the noise map in order to obtain a significance map. We find that the SW (NE) mass peak is detected at a significance level of $5.3 \sigma$ ($4.7 \sigma$).

\section{Dynamical parameters via Simulated analogs}\label{sec-analogs}

A technique recently developed by \citet{WCN18analogviewingangle} and \citet{Wittman19analogs} uses observationally derived parameters for a merger---such as subcluster mass and projected separation---to find analog systems in cosmological simulations. This allows for estimating important merger parameters that are not immediately available from observations---e.g. TSP, pericenter distance, and maximum velocity---and reconstructing the merger scenario. The analogs could also form the basis for resimulating the merger process at higher resolution and with more physics.

We select DM halo pairs undergoing merging processes in the Big Multidark Planck Simulation \citep{BigMDPL2016} and conduct mock observations by varying the viewing angle. For each analog, we compute its likelihood of matching the observed parameters of our cluster as a function of the viewing angle. The input parameters required to calculate the likelihoods are the projected separation \dproj\ between mass peaks, the line-of-sight relative velocity \dv, and the masses of the two subclusters. The analog's likelihood is then used as its weight when computing the derived dynamical parameters of the merger. For more details on the method, we refer to \citet{Wittman19analogs}.

In Section \ref{sec-wl}, we measured a projected separation of 520 kpc. Since the analog method takes symmetric error bars to calculate the likehoods, we use an uncertainty of $\pm 148$ kpc on this measurement, which represents a $\sim 70\%$ confidence interval. We also used \dv\ = $350\pm 507 \; \si{km \; s^{-1}}$ from Section \ref{sec-subclustering}, as well as the subcluster masses estimated in Section \ref{sec-wl} after accounting for our adopted value of $h$$: M_{\rm 200c}=5.2\pm 1.6\times10^{14}$ M$_\odot$ and
$M_{\rm 200c}=5.5\pm 1.6\times10^{14}$ M$_\odot$ for the SW and NE
subclusters, respectively. 

The resulting $68\%$ and $95\%$ highest probability density confidence intervals of the inferred parameters are listed in Table~\ref{tab-dynprop}. Our results suggest that RMJ1508 is a merger seen shortly after pericenter: the TSP is between 96 and 236 Myr at the $68 \%$ confidence level. For comparison, both the lower and upper boundaries of this confidence interval are lower than that of all 11 clusters in the sample considered by \citet{Wittman19analogs}. The maximum relative speed between the two subclusters ($v_{\rm max}$), however, is comparable to the median of their sample. The angle $\theta$ denotes the angle between the subcluster separation vector and the line of sight; i.e., $\theta=90^{\circ}$ is a merger entirely in the plane of the sky. Our results indicate a small line-of-sight component and are consistent with a merger in the plane of the sky. The angle $\varphi$ is defined as the angle between the current separation and velocity vectors. When $\varphi<90^{\circ}$ ($>90^{\circ}$), the merger is in its outbound (returning) phase. Also, $\varphi=0$ would indicate a fully head-on collision and a negligible impact parameter. The inferred value of $\varphi$ for RMJ1508 suggests a collision that is largely head-on. The analogs are overwhelmingly in the outbound phase: the likelihood ratio of outbound to returning analogs is 384:1.

\begin{table}
  \centering  
  \caption{Dynamical parameters from analogs}
  \begin{tabular}{ccccc}
  Confidence & TSP (Myr) & $v_{\rm max}$ ($\si{km \; s^{-1}}$) & $\theta$ (deg) &
                                                                   $\varphi$
                                                                   (deg)\\
    \hline
68\% & 96-236 &2062-2656 &62-90 &4-26\\
95\% & 0-357 &1601-2824 &36-90 &1-50
  \end{tabular}
  \label{tab-dynprop}
\end{table}

\section{Radio observations and results}\label{sec-GMRT}

Extended radio emission from galaxy clusters in the form of radio relics and halos is often a valuable resource to help constrain the merger scenario. A recent attempt at identifying extended radio sources in RMJ1508 with data from the LOFAR Two-meter Sky Survey yielded inconclusive results due to poor image quality in the cluster region \citep{Botteon2022}, encouraging new observations.

We were granted 15 hr on the upgraded Giant Metrewave Radio Telescope \citep[uGMRT,][]{uGMRT}
for Band 4 (550--900 MHz) observations of this cluster (proposal code
42\_069) with a 4\arcs\ synthesized beam size. Observations
were taken on 2022 July 4 and July 19. We used the SPAM pipeline
\citep{IntemaSPAM} to calibrate the visibilities and
\texttt{wsclean} \citep{wsclean2014,wsclean2017} to create an image with a noise level of \SI{50}{ \micro Jy \; beam^{-1}}. We found no obvious extended emission in the cluster region.

\begin{figure}
\centerline{\includegraphics[width=\columnwidth]{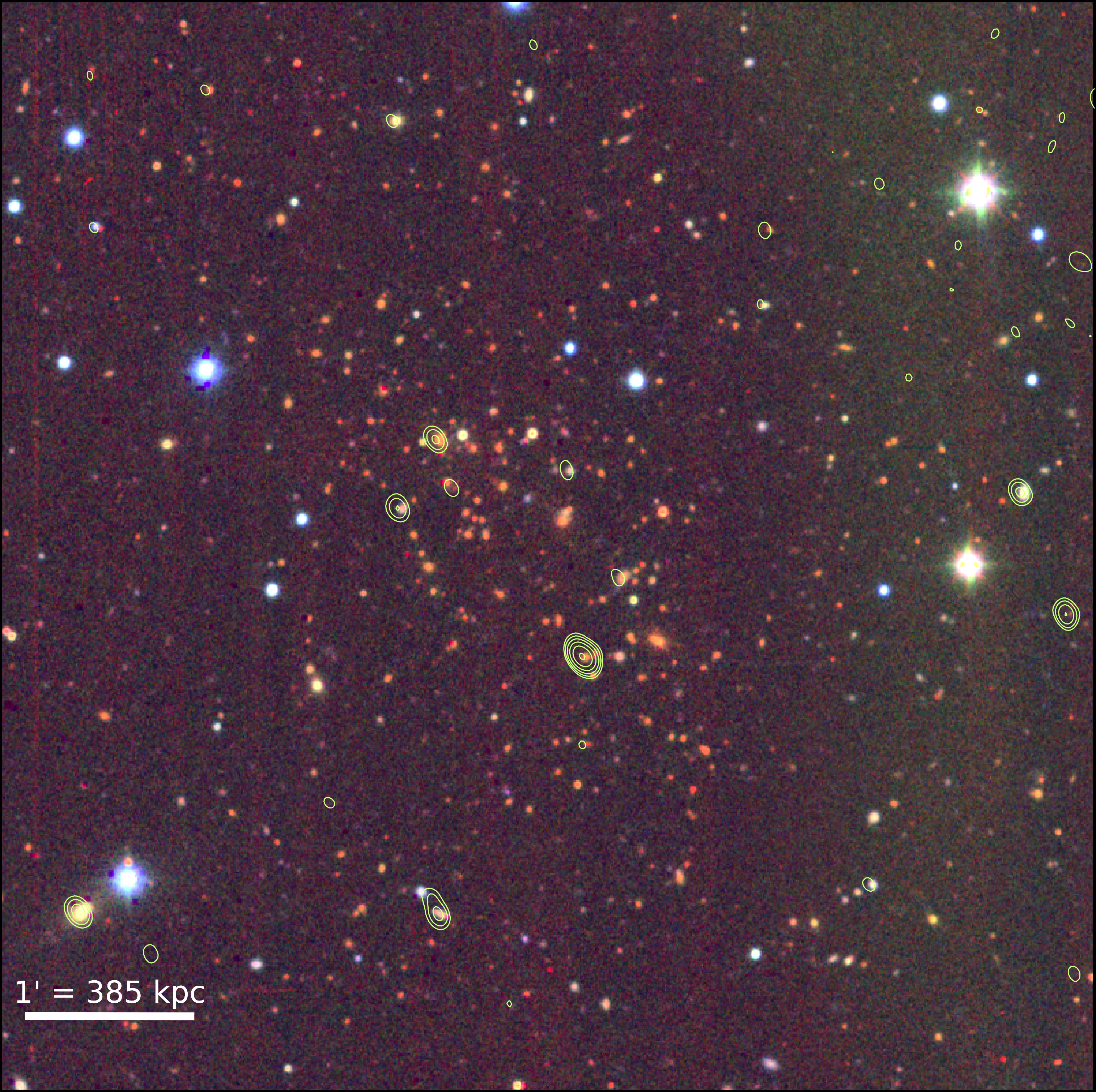}}
\caption{GMRT 650 MHz contours over a BASS multiband image. The first contour level is at \SI{80}{ \micro Jy \; beam^{-1}}, and subsequent contour levels are multiplied by a factor of 2.}
\label{fig-GMRT}
\end{figure}

Figure~\ref{fig-GMRT} shows the radio contours from the resulting image overlaid on a Beijing-Arizona Sky Survey (BASS) multiband image\footnote{Retrieved using \texttt{astroquery.hips2fits} \citep{astroquery}.} \citep{BASS}. A relatively bright point source can be seen in the SW subcluster, near the image center. Our spectroscopic data confirm that the galaxy at the origin of this radio emission is at the cluster redshift. Of the five other fainter sources in the central cluster region, four are centered on galaxies we have spectroscopic redshifts for, all of which are also at the cluster redshift.

\section{Merger scenario}\label{sec-discussion}

The X-ray morphology of RMJ1508 shows a projected distance between the two X-ray peaks in the plane of the sky of $411.6 \pm 16.3$ kpc, which is a smaller projected separation than that measured between the two BCGs (697 kpc). This projected distance is also smaller than the projected distance between the two WL peaks ($520^{+162}_{-125}$ kpc, from Section \ref{sec-wl}), although the large uncertainties in the WL measurements make this difference not statistically significant.

In the SW subcluster, the X-ray surface brightness exhibits a local peak that is removed from the corresponding BCG and lags behind the WL-peak position by a little more than $1 \sigma$ (see Figure \ref{fig-wlmap}), suggesting a dissociation between the ICM gas and DM. Assuming a postpericenter, outgoing merger scenario, the gas of the SW subcluster trails the associated DM halo and BCG as the two subclusters move away from each other toward apocenter, due to the ram pressure resulting from the interaction with the other subcluster's ICM. The NE X-ray peak, which presents a higher surface brightness than the SW peak, also lags behind its corresponding BCG, although the broader profile of the NE mass peak doesn't allow for a definitive conclusion on the DM/ICM dissociation. 
The NE BCG exhibits strong-lensing features (see Figure \ref{fig-ovw}), which could potentially be used to conduct a strong-lensing analysis and improve the constraints on the NE mass distribution. However, the redshift of the lensed galaxy remains unknown.

\begin{figure}
\centerline{\includegraphics[width=\columnwidth]{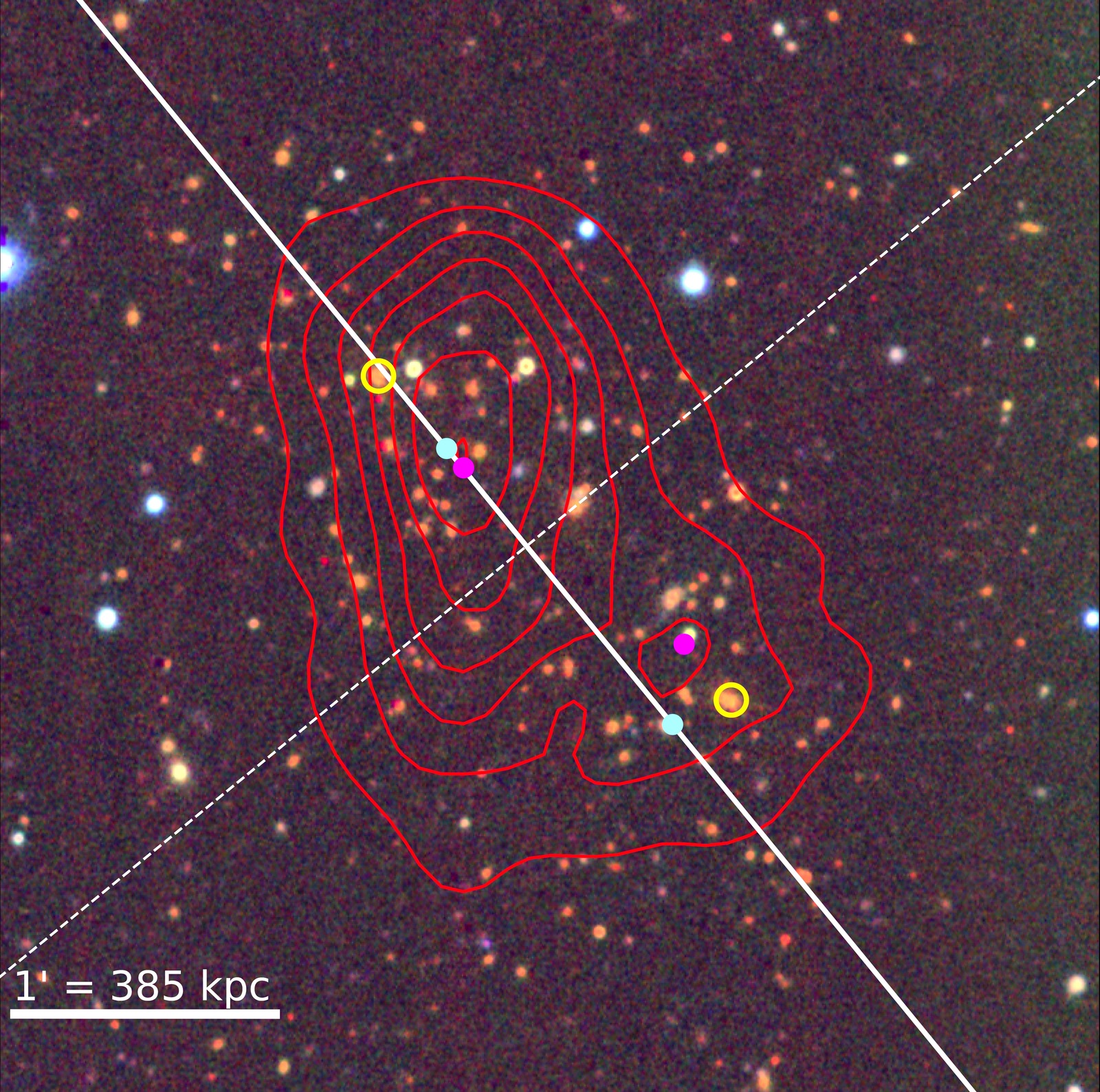}}
\caption{Definition of the merger axis (white solid line) using the centroids of the bootstrap-resampled WL peaks (cyan circles). The dashed line shows the perpendicular axis that passes through BCG-C, defining the origin of the coordinate system. The two BCGs are circled in yellow, and the X-ray surface brightness peaks are marked in magenta. 
The background image is from BASS with overlaid X-ray contours.}
\label{fig-def_axis}
\end{figure}

In order to describe the merger scenario, we define the merger axis as the line connecting the SW and NE centroids of the mass-peak positions, which are calculated from our 10,000 bootstrap realizations of the WL convergence maps (as explained in Section \ref{sec-wl}). We find the position angle of this axis to be 39$^\circ$ E of N. We define the origin of our merger axis at the projected position of BCG-C 
(circled in magenta in Figure \ref{fig-ovw}). The axis is shown in Figure \ref{fig-def_axis}, along with the positions of the BCGs, the X-ray peaks, and the WL peaks.

\begin{figure}
\centerline{\includegraphics[width=\columnwidth]{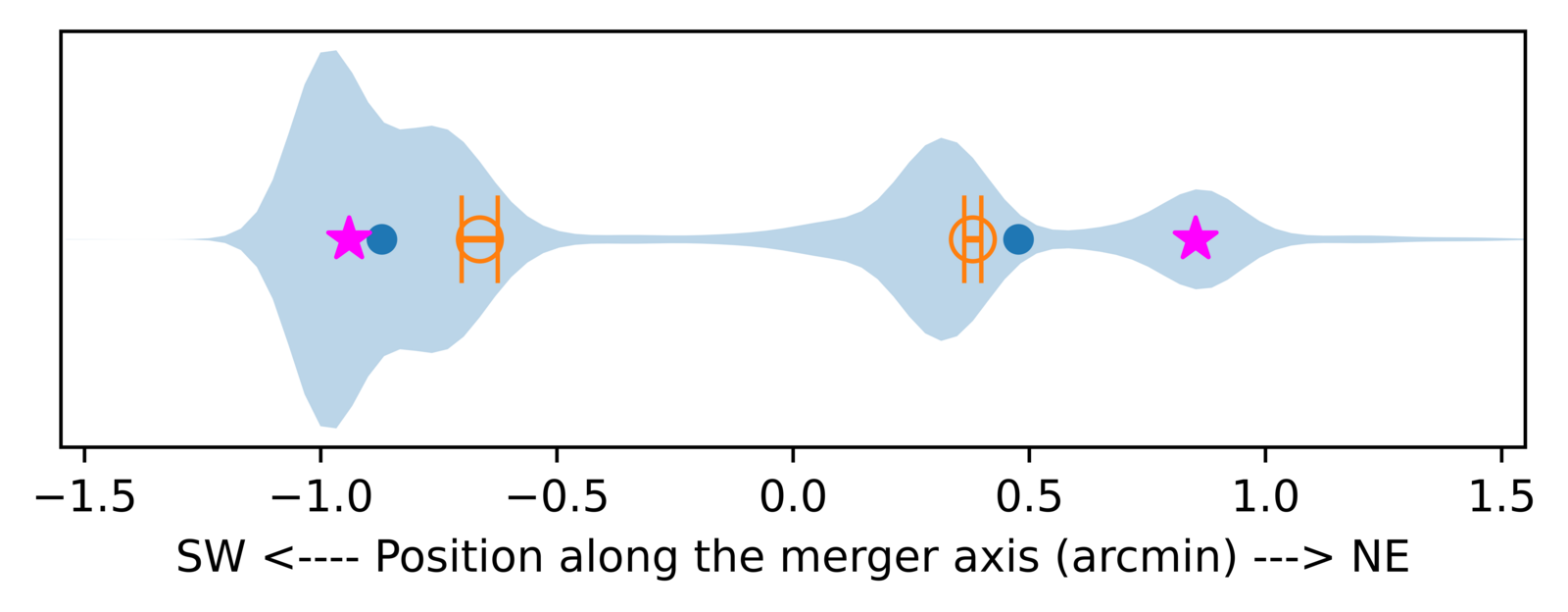}}
\caption{Subcluster components along the merger axis. The orange markers show the X-ray peaks, and the magenta stars show the BCG positions. The violin plot represents the density of bootstrap-resampled WL mass map peaks. The centroid of the WL peak distribution for each subcluster (blue circles) is computed by the k-means algorithm with two components. The error bars for the X-ray peak positions were estimated by fitting two Gaussians to the projected surface brightness profile along the merger axis and finding the uncertainty in the fitted positions.}
\label{fig-merger_scenario}
\end{figure}

The projected positions along the merger axis of the WL and X-ray peaks and of the BCG for each subcluster are shown in Figure \ref{fig-merger_scenario}. The violin plot displays the density along the merger axis of the WL mass map peaks for the 10,000 bootstrap realizations. 
Note that both the mass map in Figure \ref{fig-wlmap} and the bootstrap-resampled WL peak distribution shown in Figure \ref{fig-merger_scenario} present a bimodal shape in the NE subcluster. For more than half of the bootstrap realizations, the NE WL peak is actually closer to the origin of the merger axis than the X-ray peak, which contradicts the expectation for a merger seen shortly after pericenter. This raises the question of whether the NE DM halo really possesses non-trivial substructure, or if this feature is an artifact of the limited sample of background galaxies. 
The answer possibly requires deeper imaging covering a larger field of view.

\begin{deluxetable}{lcc}
  \tablecaption{Distance Along Merger Axis (kpc)}  \label{tab-distances}
  \tablecolumns{3}
  \tablehead{ & \colhead{NE} & \colhead{SW} }
  \startdata
    WL -- BCG &  $-144.7^{+147.6}_{-92.6}$ & $26.7^{+60.2}_{-50.4}$  \\
    X-ray -- BCG & $-182.0 \pm 6.9$ & $106.6 \pm 14.7$  \\
    WL -- X-ray &  $37.3^{+147.8}_{-92.9}$ & $-79.9^{+61.9}_{-52.5}$ 
    \enddata
\end{deluxetable}

The distances along the merger axis between all three measured components of each subcluster are listed in Table \ref{tab-distances}. For the SW subcluster, the projected separation between the WL and X-ray peaks is larger than the error bars, suggesting that the peak of the DM distribution is 
likely 
to be located to the south of the gas density peak. Although there is a $\sim$30 $\; \si{kpc}$ distance between the projected positions of the mass peak and the BCG, this distance is well within the limits of the WL uncertainty. For the NE subcluster, the X-ray peak is located $\sim$180 $\; \si{kpc}$ to the SW of the BCG. However, the previously discussed morphology of the NE mass profile results in large uncertainties in the position of its peak along the merger axis, prohibiting a clear determination of the relative positions between the WL and X-ray peaks, as well as the WL peak and BCG.

\section{Summary}\label{sec-summary}

Our multiwavelength analysis 
suggests 
that RMJ1508 is a cluster in a merging state seen shortly after first pericenter passage. By fitting a two-halo Navarro–Frenk–White (NFW) model to the WL mass maps, we infer that the two subclusters have approximately the same mass. The NE subcluster, however, has a broader mass peak than the SW, which is somewhat more well defined. The X-ray morphology, revealed by \XMM\ archival data, exhibits two surface brightness peaks between the subclusters' BCGs, in a classic dissociative merger configuration.

Our spectroscopy results reveal that the line-of-sight component of the relative velocities between the two subclusters is small, suggesting a merger very close to the plane of the sky. By selecting analog systems in cosmological simulations, we confirm that the merging axis is $\leq 28^{\circ}$ from the plane of the sky at the $68\%$ confidence level.  The analogs also indicate a TSP ranging from 96 to 236 Myr.

In summary, the system we have presented possesses many features that make it a great candidate for further study. The two equal-mass subclusters are colliding head-on, resulting in a likely dissociation between DM and ICM components, especially in the SW subcluster. The merger axis is conveniently placed near the plane of the sky, and the cluster's simple, binary nature makes the merger scenario relatively easy to identify and model.

Follow-up observations of RMJ1508 could improve our understanding of the system. For example, ground-based imaging suitable for WL analysis could help improve the accuracy of the mass maps and either confirm or rule out the substructure observed in the NE mass peak. In addition, the observed merger parameters could be used to stage high-resolution hydrosimulations and leverage the system's suitability to test DM models, potentially improving the constraints on the DM self-interaction cross section. Moreover, our merger-finding method, which enabled the discovery of this cluster, is well suited to take advantage of upcoming optical and X-ray surveys such as the Legacy Survey of
Space and Time \citep{LSST2019} and e-ROSITA and has the potential to significantly expand the ensemble of binary, dissociative merging clusters that can help uncover new physics.

\acknowledgments

We thank the anonymous referee for the valuable comments and suggestions. We thank Gastão Lima Neto for guidance in the \XMM\ data reduction. We thank Nissim Kanekar for help with GMRT exposure time calculations, and Huib Intema for help with the SPAM pipeline. We thank Reinout van Weeren and Andrea Botteon for guidance in the GMRT data reduction. 
This work was supported by NSF grant No.
2308383. We thank the staff of the GMRT that made these observations possible. GMRT is run by the National Centre for Radio Astrophysics of the Tata Institute of Fundamental Research. Some of the data presented herein were obtained at Keck Observatory, which is a private 501(c)3 nonprofit organization operated as a scientific partnership among the California Institute of Technology, the University of California, and the National Aeronautics and Space Administration. The Observatory was made possible by the generous financial support of the W. M. Keck Foundation. The authors wish to recognize and acknowledge the very significant cultural role and reverence that the summit of Maunakea has always had within the Native Hawaiian community. We are most fortunate to have the opportunity to conduct observations from this mountain.
Some of the data presented in this paper were
obtained from the Mikulski Archive for Space Telescopes (MAST) at the
Space Telescope Science Institute. 
The specific observations analyzed can be accessed via \dataset[10.17909/bs78-sw95]{https://doi.org/10.17909/bs78-sw95}.

\facilities{Keck:II (Deimos), HST (ACS), GMRT, XMM} 

\software{SAS (v19.0.0; \citet{SAS}), XSpec (v12.11.1; \citet{XSPEC}), mc3gmm code \citep{MCCsampleanalysis}, FIATMAP code \citep{2006wittmanDLS}, SExtractor \citep{1996bertin}, GalSim \citep{galsim}, SciPy \citep{scipy}, Astropy \citep{astropy2022}, Matplotlib \citep{matplotlib}, NumPy \citep{numpy}, Astroquery \citep{astroquery}, Scikit-learn \citep{sklearn}}
  
\vskip10pt
\bibliography{ms}
\bibliographystyle{aasjournal}

\end{document}

%% file: ztable.tex
226.90117 & 57.83078 & 0.33552 & 0.00010\\ 
226.90813 & 57.87533 & 0.38313 & 0.00011\\ 
226.92946 & 57.82292 & 0.61105 & 0.00023\\ 
226.93522 & 57.89799 & 0.35119 & 0.00011\\ 
226.93679 & 57.82869 & 0.60394 & 0.00011\\ 
226.94302 & 57.83171 & 0.42396 & 0.00010\\ 
226.94401 & 57.88695 & 0.35363 & 0.00010\\ 
226.94578 & 57.86279 & 0.53800 & 0.00011\\ 
226.94750 & 57.88132 & 0.35382 & 0.00010\\ 
226.95028 & 57.83540 & 0.39716 & 0.00012\\ 